\documentclass[12pt]{article}
\begin{document}
\title{The Life and Legacy of Pomeranchuk}
\author{L.B. OKUN
\thanks{ITEP, Moscow, Russia 
E-mail: okun@heron.itep.ru}
}
\date{}
\maketitle
\section{Introduction}

The life of Isaak Yakovlevich Pomeranchuk was short (20.05.1913 --
14.12.1966). But the impact of his personality and his works on
physics and physicists is remarkable.

With better luck he could live today and even participate in this
conference. It is difficult to imagine what would be the theme of
his talk. The 36 years without him brought drastic changes in all
aspects of life, science in general and physics in particular. But
many of the talks at this conference have roots in the ideas and
the thrust of Pomeranchuk.

All his life he was ready to start a new journey into the world of
unknown. He went on many of them. Insatiable  scientific curiosity
was one of his dominant passions.

\section{On the way to physics}

Yuzik Pomeranchuk was born in Warsaw (at that time in Russia) on
the eve of the World War I. His father was a chemical engineer,
mother -- medical doctor. In 1918 the family moved to
Rostov-on-Don, later, in 1923, -- to a town Rubezhnoe in Donets
Basin, where his father worked as engineer at a chemical plant. In
1927 Yuzik graduated from 7-year school and two years later from a
factory - and workshop school. In 1929-31 he was a worker at
chemical plant. In 1931 he left Rubezhnoe for Ivanovo, a city some
300 km North-East of Moscow, and became a first year student of
the Institute of Chemical Technology. In 1932 he moved from
Ivanovo Institute to the second course of Physical-Mechanical
Department of the Leningrad Polytechnical Institute. Here he
specialized in chemical physics. Fellow students recalled that
Yuzik enthusiastically worked 14 hours a day and led rather
ascetic life. He searched in the library answers to questions
which appeared during lectures and had a special gift of clearly
explaining to his friends all difficult problems.

\section{Yuzik, are you a theorist?}

In 1934-35 his advisor became Alexander Iosifovich Shalnikov (1905
-- 1986). Many years later academician Shalnikov recalled that he
started by bringing Yuzik to a room full with old vacuum pumps
glass-ware and leaving him there. Two weeks later, entering the
room, Shalnikov found that everything that could be broken was
broken. `Yuzik, are you a theorist?'', -- uttered Shalnikov. ``I
don't know. And what?'' -- was the answer.\footnote{This and
future quotations are from the book ``Vospominaniya o
I.Ya.~Pomeranchuke'', Moskva, Nauka, 1988 (in Russian), which
contains recollections of 83 authors and 7 scientific reviews:
describing the impact of Pomeranchuk on physics of elementary
particles, quantum field theory, theory of nuclear reactors,
synchrotron radiation, quantum liquids and crystals, heat
conductivity in dielectrics.}

Shalnikov and his wife bought a ticket and put Yuzik on a train to
Kharkov, providing him with some money and food (it was a lean
year). Kharkov was known at that time as a new center of
theoretical physics led by the young Lev Davidovich Landau (1908
-- 1968), who spent  before a few years at the Niels Bohr
Institute in Copenhagen and decided to create a similar cradle for
young theorists in the Soviet Union, Yuzik became one of the first
students of Landau. He passed the famous ``theoretical minimum
exams'' in two months (for many others it took years) and
published his first paper on light by light scattering, together
with A.~Akhiezer and L.~Landau in 1936 in ``Nature''.

Pomeranchuk was a devoted disciple of Landau. (The names Dau and
Chuk were coined by Landau.) ``I am ready to go after Dau to the
Dickson island'', said young Chuk to one of his friends. (Dickson
is an island in the Arctic Ocean near the shores of Siberia, at
the mouth of Yenisey.) The mutual respect and affection thread
runs all through their lives.

\section{The 1936-1941 Firework}

The ``photon-photon'' article started a firework of papers in
various fields, such as cosmic rays, neutron scattering in
crystals, conduction of heat and sound in dielectrics,
electroconductivity of metals, superconductivity and other
subjects.

The most unexpected was the discovery that there should exist an
upper limit about $10^{17}$ eV on cosmic electron energy at the
surface of the Earth due to emission of photons in the magnetic
field of the Earth. Arkady Migdal recalled that when Pomeranchuk
told him about his idea, he (Migdal) was convinced that such a
weak field could not influence the trajectory of ultra-high energy
particle. But intuition of Pomeranchuk turned out to be correct.
This effect was the first in a series of famous works of
Pomeranchuk on the theory of synchrotron radiation.

Among other results in condensed matter physics was the correction
to Peierls $1/T$ law for heat conduction at high temperature $T$.
``It was not pleasant, but it consoled me that I had been caught
by very clever Pomeranchuk but not by somebody else'', recalled
Rudolf Peierls.

Not everything, on which Pomeranchuk worked at that time, was
published by him. In 1941 he gave three talks at a conference held
in Moscow. The program of the conference was published in
``Journal of Physics of USSR''. Two talks by Pomeranchuk ``The
production of mesotron pairs by positron annihilation'' and `The
scattering of mesotrons by mesotrons'' are presented by the
abstracts, while the third one ``Nuclear reactions inside stars''
is presented by the title only.

\section{Wanderings: 1937 -- 1946}

In 1937 under the threat of arrest Landau left Kharkov for Moscow,
to join the Kapitza Institute. Pomeranchuk followed him to become
an assistant lecturer at the Moscow institute of tanning industry.

The transfer to Moscow delayed Landau's arrest only by a year: in
1938 he was imprisoned. Only a letter by Peter Kapitza to Stalin
saved Landau's life, he spent in the prison a year.

After Landau's arrest Pomeranchuk moved to Leningrad. In 1938-39
he is an assistant lecturer at Leningrad University, defending
there in 1938 his PhD. In 1939-40 Pomeranchuk is a junior
scientist at Leningrad Physical-Technical Institute. In 1940 he
comes back to Moscow as a senior scientist at Lebedev Institute
and defends his DSc dissertation ``Heat conductivity and
absorption of sound in dielectrics''. In 1941 the war started and
Lebedev Institute was evacuated to Kazan. From there in 1942
Pomeranchuk was sent as a member of a group headed by Abram
Isaakovich Alikhanov to Armenia. The task of the group was to
initiate the construction of a station for study of cosmic rays at
the Aragats mountain. They were on their way to Armenia during the
Stalingrad battle. In Yerevan Pomeranchuk together with
A.F.~Kirpichev wrote three papers on the theory of cosmic ray
showers.

In May 1943 Pomeranchuk returned to Moscow to become a member of
the Kurchatov team, working on the first Soviet nuclear reactor.
He worked with Igor Vasilyevich Kurchatov at what was called Lab.
No.2 and is now called Kurchatov Institute till the beginning of
1946. For the rest of his life he stayed at what then was called
Lab. No.3 and is now called ITEP being the founder and head of the
ITEP theory division. Also in 1946 Pomeranchuk became Professor of
theoretical physics at Moscow Mechanical Institute (at present
MEPHI).

\section{Neutrons in 1940s}

Since 1943 Pomeranchuk was the leader of nuclear reactor theory in
Soviet Union. He developed the theory of exponential experiments
which allowed Kurchatov to measure the neutron absorption in
moderator and the value of moderation length.

Neutrons became dominant in his work in 1940s. Together with
I.~Gurevich he created a theory of resonance absorption of
neutrons in heterogeneous nuclear reactor with natural uranium.
The construction of the first Soviet reactor completed in 1946 was
based on this theory. (It was published as a report at the
International conference on peaceful applications of atomic energy
in 1955.) The theory of neutron absorption in homogeneous media
was developed by Pomeranchuk and Akhiezer in the manuscript
``Introduction to the theory of neutron multiplying systems'',
ITEP, 1947. It served as a basic manual for construction of Soviet
nuclear reactors for many years. I remember using it as a novice
calculating  nuclear reactor for China in the middle of 1950s. At
that time its single copy was worn out to pieces. In 2002 the book
was ultimately edited and published by B.~Ioffe and A.~Gerasimov.
A part of this manuscript was published as a separate book:
A.~Akhiezer and I.~Pomeranchuk ``Certain problems of nuclear
theory'' in 1948 and then in 1950. Theory of nuclear reactors was
the main practical task of the theory division created at ITEP by
Pomeranchuk  who recruited V.~Berestetsky, A.~Galanin, B.~Ioffe,
A.~Rudik. The non-reactor physics was considered as a kind of a
hobby.

\section{``Hobby problems'' in 1940s}

The number, variety and quality of these ``hobby problems'' is
quite impressive: Pomeranchuk continued his research on phonons
(heat and sound conduction), on cosmic rays (on the atmospheric
showers), on spectrum of synchrotron radiation at electron
accelerators. He published two articles: on maximal energy of
electrons in a betatron (1944, together with D.~Ivanenko) and on
spectrum of synchrotron radiation (1946, together with
L.~Artsimovich). By the way, V.~Vladimirsky found in 1948 that
maximal energy of showers produced by cosmic electrons in the
atmosphere can be much larger than the Pomeranchuk limit for a
primary electron, because above $10^{18}$ eV the quantum of
radiation carries substantial part of electron's energy.
Pomeranchuk helped him to publish this result.

New themes appeared:

The first theme: liquid helium, in particular helium-3. In 1948 he
established that both $^3$He and $^6$He admixtures to $^4$He
belong to normal (non-superfluid) component of liquid helium. This
research culminated by the famous 1950 paper in which Pomeranchuk
has put forward the idea that the entropy of helium-3 is smaller
in liquid state than in the solid one and suggested his method of
reaching microkelvin temperatures. (The nuclear spins of $^3$He
should be aligned in the liquid phase (due to exchange forces) and
not aligned in the solid phase. Hence entropy of the latter is
larger.)

Implementation and development of these ideas were crowned in 1996
by Nobel Prize to David Lee, Douglas Osheroff and Robert
Richardson for their discovery of superfluidity in helium-3. We
have the privilege to listen today to the talk of Professor
Osheroff.

The second theme: 1945, the finite size atomic nuclei with $Z >
137$. Y.A.~Smorodinsky who coauthored this article recalled that
Pomeranchuk's enthusiastic exclamation ``It would be great to
collide two uranium nuclei!'' was met by laughter: nobody believed
in such experiments. They were done much later in Darmstadt.

The third theme: 1948, selection rules in the decay of positronium
(the three photon decay of orthopositronium and two photon decay
of parapositronium).

\section{Arzamas-16}

Early in 1950 Pomeranchuk was sent to the atomic weapon center at
Arzamas-16. The order was signed by Stalin. Pomeranchuk longed for
his family (wife and his stepdaughters) and for ``hobby physics''.
He used to come to N.~Bogolubov (they lived in the same hostel) to
discuss with him how to revoke the order. Bogolubov entertained
him with a cup of excellent strong coffee and aphoristically
uttered: ``Orders are not revoked, usually they are forgotten''.
Within a year Pomeranchuk was again in Moscow, at ITEP.

\section{QFT and zero charge}

Pomeranchuk enthusiastically reacted to the breakthrough in
quantum field theory. In 1951 he set up a special seminar to study
the relativistic covariant technique originated by Feynman, Dyson,
Schwinger and others. He used it in two different ways: to
elucidate the basic problems of the renormalization procedure and
to apply it to various specific phenomena.

On the first way he and Landau discovered the fundamental
difficulty which they dubbed ``zero charge'', and which is now
referred to as ``triviality problem''. They found that any finite
value of ``bare charge'' (at short distances) results in a
vanishing ``physical charge'', i.e. in the absence of interaction
(at observable distances). This property was established first in
quantum electrodynamics, then in various types of meson theories.
An important role at this stage was played by V.~Sudakov and
K.~Ter-Martirosyan who joined ITEP in the 1950's.

Only in the early 1970's it became clear that the non-abelian
gauge theories do not suffer from this fatal disease. As you go to
shorter distances the non-abelian charges decrease providing
``asymptotic freedom''. But even today the disease discovered by
Pomeranchuk and Landau is not eradicated. It is hiding in the
Higgs sector of the electroweak interaction. The fight against
this disease is one of the driving forces of the modern field
theory.

\section{Applications of QFT and $S$-matrix}

In spite of unsolved ``zero charge'' basic inconsistency
Pomeranchuk vigorously applied QFT to various phenomena which did
not ``feel'' this inconsistency. Together with A.~Galanin in 1952
he calculated the energy shift in muonic atoms due to polarization
of vacuum by virtual photon. He modified in 1951 the approach
suggested by E.~Fermi to multiple production of pions at high
energies.

Pomeranchuk paid great attention to the characteristic time
intervals at which various processes took place. In 1953 together
with L.~Landau he found how the standard Bethe-Heitler formula for
Bremsstrahlung cross section is modified in a medium. The
characteristic time of the process is $E^2/m^2 \omega$, where $E$
-- energy of the electron, $m$ -- its mass, and $\omega$ --
frequency of the photon. At high enough energy the emission length
becomes so large that multiple scattering is essential. Further
refinement of the problem was done by A.~Migdal. The LPM-effect is
under active study even at present.

In 1953 Pomeranchuk was elected corresponding member of Soviet
Academy. He became full member (academician) in 1964.

In 1950's Pomeranchuk analyzed diffraction processes in the
collisions of nucleons at high energies and peripheral collisions
with one-pion and two-pion exchange. Here he relied on dispersion
relations, which follow from such general principles as causality
and conservation of probability (analiticity and unitarity  of
$S$-matrix).

On the basis of dispersion relations and of assumption that the
radius of a particle does not increase with its energy he proved
in 1958 his famous theorem, according to which the cross-sections
of a particle and its antiparticle on a given target are equal to
each other at asymptotically high energy.

Pomeranchuk theorem gave a strong impetus to experiments and
theory in particle physics, to building high energy accelerators
and colliders. In 1958 the highest energy was about 1 GeV. Today
it is measured by TeVs.

\section{Pomeron}

Theoretical analysis has shown quite soon that the assumption that
the radius of, say, proton does not depend on energy, is too
naive. It should increase as logarithm of energy. But asymptotic
equality of particle and antiparticle cross-sections discovered by
Pomeranchuk remained valid.

The asymptotic behaviour in $s$-channel depends on the angular
momentum $j(t)$ of quasi-particles exchanged in $t$-channel ($s=
E^2$, $t=q^2$): $s^{j(t)}$. This angular momentum depends on $t$,
forming in the plane $j, t$ the so-called Regge-trajectory.
Hadrons with integer spin (mesons) and half-integer spin (baryons)
lie on these trajectories.

The behaviour of total cross-section depends on the value of $j$
at $t=0$ of the trajectory with all quantum numbers (charge,
isospin, $C$-parity) of the vacuum. It was M.~Gell-Mann who
suggested to call this vacuum trajectory the Pomeranchuk
trajectory. The corresponding quasi-particle is referred to as
pomeron in the literature.

Pomeranchuk published about a dozen articles on reggeonic theory
of strong interactions (most of them with Vladimir Naumovich
Gribov (1930-1997)). The first of these articles was published in
1962, the last -- after Pomeranchuk passed away.

During last four years of Pomeranchuk's life his collaboration
with Gribov was fantastically intense. Gribov, who lived in
Leningrad, regularly came to ITEP, a few times Pomeranchuk took
train to Leningrad. Their discussions in Pomeranchuk's office at
ITEP lasted late into the night. Both of them were heavy smokers.
A cigarette was lighted from the previous one. The smoke in the
room was thick. Many years later J.~Bj\"{o}rken wrote that those
brilliant works had cleared the mist in the minds of theorists at
Berkeley and brightly demonstrated the extremely high level of
research in certain fields in the Soviet Union.

\section{His disciples}

Pomeranchuk influenced the development of physics not only through
his articles. He was a Great Teacher. Being the head of theory
division at ITEP, he at the same time was a Professor of
theoretical physics at Moscow Engineering Pysics Institute
(MEPHI), which is hosting this memorial conference. His former
students were A.~Rudik, I.~Kobzarev, M.~Terentyev, Yu.~Nikitin,
A.~Kaidalov, Yu.~Simonov,   Yu.~Vdovin, I.~Dremin, V.~Mur,
 E.~Zhizhin, A.~Berkov, B.~Karnakov, I.~Zuckerman and among many
others myself.

When in 1951 Pomeranchuk was designated to become the head of the
theory group in Dubna he brought with him one of his MEPHI
students S.~Bilenky. For about a year Pomeranchuk used to come to
Dubna every week as a part time job. Among his disciples there was
L.~Lapidus. Pomeranchuk strongly influenced the work of R.~Ryndin
and of many experimentalists. According to recollections of
B.~Pontecorvo, V.~Dzhelepov and many others each visit by
Pomeranchuk was a remarkable event in the life of physicists in
Dubna, who remembered his improvised seminars for many years.

ITEP theory division  attracted also students from Moscow
University. The most prominent of them are B.~Ioffe, M.~Marinov
and V.~Popov.

The personality of Pomeranchuk is unforgettable to everybody who
had the luck to work with him, to listen to his lectures, or even
only to meet him. He was extremely devoted to physics and at the
same time he was very human. Great integrity was fused in him with
sincere sympathy, with rare sense of humor. Many of his jokes help
us at ITEP to live through hard times.

\section{Fight against cancer}

Pomeranchuk did not look an athlete, but he was strong, of great
endurance and of great courage. In the fall of 1965 doctors
discovered that he was fatally ill: cancer of esophagus. For more
than a year he underwent chemotherapy, radiotherapy, surgery
(gastrotomy). But he continued to do physics in hospitals and at
home till his last day.

During the tragic days and nights of his illness Pomeranchuk was
cared for by his stepdaughter Marina who fought for his life with
inexhaustible energy. Today Marina Alexeevna Ivanova-Pomeranchuk
is with us in this hall.

Being treated by gamma-rays he thought about the Bragg peak in
stopping proton beam. It was obvious for him that protons, unlike
electrons or gamma-rays, would deliver most of their energy at the
end of their track inside the tumor. He summoned physicists from
ITEP and Dubna together with radiologist in the Oncological Center
to start the program of proton therapy at the existing proton
accelerators. Unfortunately it was too late to help him. But he
thought about others. The first medical proton beam started to
operate at ITEP in 1969. More than 3000 patients were successfully
treated at ITEP. Recently special centers for proton therapy were
built in Western Europe and USA.

\section{Pomeranchuk Prize}

In 1998 Pomeranchuk Prize was established at ITEP. Up to now the
international jury awarded it to ten physicists:

A.~Akhiezer, S.~Drell (1998)

K.~Ter-Martirosyan, G.~Veneziano (1999)

E.~Feinberg, J.~Bjorken (2000)

L.~Lipatov, T.~Regge (2001)

L.~Faddeev, B.~DeWitt (2002)

Professor B.~DeWitt will receive the Prize today. Professor
L.~Faddeev -- at the ITEP International Winter School in February.

The names of 2003 prize winners will be announced on 20th of May
2003, Pomeranchuk's 90th birthday.

The constellation of laureates sheds a special light on the legacy
of Pomeranchuk: on his formulas and ideas in various fields of
physics, on his style of research marked with deep intuition,
ingenuity and hard work, on his devotion to physics, his
integrity, his striving to the heart of the matter, combined with
painstaking attention to details, his gift of a great teacher who
shared his enthusiasm and knowledge with his disciples and peers.

\section{Addendum}

\begin{center}

{\bf List of scientific publications by I.Ya.
Pomeranchuk}\footnote{This is a translation into English of a list
published in the third volume of Scientific Works (SW) by
I.Ya.~Pomeranchuk, ``Nauka'', Moskva, 1972 (in Russian).}

\vspace{5mm}

1936 -- 1941

\vspace{3mm}

\end{center}

\begin{enumerate}
\item
On properties of metals at very low energies. (With L.D.~Landau.)
ZhETF {\bf 7} (1937) 379; Phys. Zs. Sowjet. {\bf 10} (1936) 649.
(SW, {\bf 1}, 1).\footnote{This is a reference to the volume and
the number of the article in SW.}
\item Light by light scattering. (With A.I.
Akhiezer and L.D. Landau.) Nature {\bf 138} (1936) 206 (SW, {\bf
2}, 33).
\item Coherent scattering of $\gamma$-rays by nuclei. (With A.I.
Akhiezer.) ZhETF {\bf 7} (1937) 567; Phys. Zs. Sowjet. {\bf 11}
(1937) 478 (SW, {\bf 2}, 34).
\item On the scattering of slow neutrons in a crystal lattice. ZhETF
{\bf 8} (1938) 894; Phys. Zs. Sowjet. {\bf 13} (1938) 65 (SW, {\bf
1}, 25).
\item Critical field in superconductors of small size. ZhETF {\bf
8} (1938) 1096 (SW, {\bf 1}, 2).
\item
On the maximal energy which the primary electrons of cosmic rays
can have on the earth's surface due to radiation in the earth's
magnetic field. ZhETF {\bf 9} (1939) 915; J. Phys. SSSR {\bf 2 }
(1940) 65 (SW, {\bf 2}, 41).
\item The influence of a magnetic field on the
electrical conductivity of bismuth single crystals at low
temperatures. (With B.I.~Davydov.) ZhETF {\bf 9} (1939) 1294; J.
Phys. USSR {\bf 2} (1940) 147 (SW, {\bf 1}, 3).
\item On the end of mesotron track in Wilson chamber. (With
A.B.~Migdal.) Dokl. AN SSSR {\bf 27} (1940) 652; Phys. Rev. {\bf
57} (1940) 934 (SW, {\bf 2}, 42).
\item The thermal conductivity of the paramagnetic dielectrics at low
temperatures. ZhETF {\bf 11} (1941) 226; J. Phys. USSR {\bf 4}
(1941) 357 (SW, {\bf 1}, 7).
\item On the thermal conductivity of dielectrics at the temperatures
higher than the Debye temperature. ZhETF {\bf 11} (1941) 246; J.
Phys. USSR {\bf 4} (1941) 259 (SW, {\bf 1}, 8).
\item Sound absorption in dielectrics. ZhETF {\bf 11} (1941)
455; J. Phys. USSR {\bf 4} (1941) 529 (SW, {\bf 2}, 9).
\item The production of mesotron pairs by positron annihilation. J. Phys.
USSR {\bf 4} (1941) 277.
\item The scattering of mesotrons by mesotrons. J. Phys. USSR {\bf 4} (1941) 277.
\item Nuclear reactions inside stars. J. Phys. USSR {\bf 4} (1941)
285.
\item On conduction of heat in dielectrics. Phys. Rev. {\bf 60}
(1941) 820 (SW, {\bf 1}, 10).

\begin{center}

1942 -- 1946

\end{center}

\item On conduction of heat in dielectrics below Debye
temperature. ZhETF {\bf 12} (1942) 245; J. Phys. USSR {\bf 6}
(1942) 237; Izvestiya AN SSSR, physical series {\bf 6} (1942) 77
(SW, {\bf 1}, 11).
\item Conduction of heat in dielectrics at high temperatures.
ZhETF {\bf 12} (1942) 419; J. Phys. USSR {\bf 7} (1943) 197 (SW,
{\bf 1}, 12).
\item Dependence of sound absorption in dielectrics on frequency
and temperature. J. Phys. USSR {\bf 7} (1943) 266 (SW, {\bf 1},
13).
\item Coulomb forces and neutron structure. Dokl. AN SSSR {\bf 41}
(1943) 162 (SW, {\bf 2}, 54).
\item Spectrum of soft component in the air at high energies.
(With A.~Kirpichev.) Dokl. AN SSSR {\bf 41} (1943) 19 (SW, {\bf
2}, 43).
\item On the theory of transition effect in cosmic rays. (With
A.~Kirpichev.) Dokl. AN SSSR {\bf 42} (1944) 396 (SW, {\bf 2},
44).
\item On the theory of absorption of infrared rays in
center-symmetrical crystals. ZhETF {\bf 13} (1943) 428; J. Phys.
USSR {\bf 7} (1943) 262 (SW, {\bf 1}, 14).
\item On the screening of effective cross sections of
Bremsstrahlung and pair production calculated with experimental
data on atomic form factors. (With A.~Kirpichev.) Dokl. AN SSSR
{\bf 45} (1944) 301 (SW, {\bf 2}, 45).
\item Scattering of mesons strongly interacting with nucleons.
Dokl. AN SSSR {\bf 44} (1944) 13 (SW, {\bf 2}, 55).
\item On the conduction of heat in magnetically cooled salts.
(With A.I.~Akhiezer.)  J. Phys. USSR {\bf 8} (1944) 216 (SW, {\bf
1}, 15).
\item On thermal equilibrium of spins and lattice. (With
A.I.~Akhiezer.) ZhETF {\bf 14} (1944) 342; J. Phys. USSR {\bf 8}
(1944) 206 (SW, {\bf 1}, 16).
\item On the interpretation of experimental data on large
avalanche showers. ZhETF {\bf 14} (1944) 252; J. Phys. USSR {\bf
8} (1944) 17 (SW, {\bf 2}, 46).
\item Radiation of fast electrons in magnetic field. (With
L.A.~Artsimovich.) ZhETF {\bf 16} (1946) 379; J. Phys. USSR {\bf
9} (1945) 267 (SW, {\bf 2}, 48).
\item On the maximum energy attainable in a betatron. (With
D.D.~Ivanenko.) Dokl. AN SSSR {\bf 44} (1944) 343 (SW, {\bf 2},
47).
\item On energy levels in systems with $Z>137$. (With
Ya.A.~Smorodinsky.) J. Phys. USSR {\bf 9} (1945) 97 (SW, {\bf 2},
35).
\item On the conduction of heat in bismuth. (With A.I.~Akhiezer.)
ZhETF {\bf 15} (1945) 587; J. Phys. USSR {\bf 9} (1945) 93 (SW,
{\bf 1}, 4).
\item On elastic scattering of fast charged particles by nuclei.
(With A.I.~Akhiezer.) ZhETF {\bf 16} (1946) 396; J. Phys. USSR
{\bf 9} (1945) 471 (SW, {\bf 3}, 96).
\item On elastic scattering of neutrons with energy of a few Kelvins
in liquid helium II. (With A.I.~Akhiezer.) ZhETF {\bf 16} (1946)
391; J. Phys. USSR {\bf 9} (1945) 461 (SW, {\bf 1}, 18).
\item Radiation of relativistic electrons in magnetic field. Izv.
AN SSSR, physical series {\bf 10} (1946) 316 (SW, {\bf 2}, 49).

\begin{center}

1947 -- 1951

\end{center}

\item On scattering of slow neutrons in crystals. (With
A.I.~Akhiezer.)  ZhETF {\bf 17} (1947) 769; J. Phys. USSR {\bf 11}
(1947) 167 (SW, {\bf 1}, 28).
\item Generalization of the limiting $\lambda$-process and
ambiguity of elimination of infinities in quantum theory of
elementary particles.  ZhETF {\bf 17} (1947) 667; J. Phys. USSR
{\bf 76} (1949) 298 (SW, {\bf 2}, 56).
\item On the theory of resonance scattering of particles. (With
A.I.~Akhiezer.)  ZhETF {\bf 18} (1948) 603 (SW, {\bf 1}, 29).
\item On refraction of neutrons. (With A.I.~Akhiezer.)  ZhETF {\bf 18} (1948)
475 (SW, {\bf 1}, 30).
\item On the motion of foreign particles in helium II. (With
L.D.~Landau.) Dokl. AN SSSR {\bf 59} (1948) 669 (SW, {\bf 1}, 19).
\item On fluctuation of ionization ranges.  ZhETF {\bf 18} (1948)
759 (SW, {\bf 2}, 50).
\item Introduction into the theory of neutron multiplicating
systems (reactors) (With A.I.~Akhiezer.) ITEP Report, M., 1947;
Moscow, Izd.AT, 2002.
\item Selection rules for annihilation of electrons and positrons.
Dokl. AN SSSR {\bf 60} (1948) 213 (SW, {\bf 2}, 36).
\item Certain problems of nuclear theory. (With A.I.~Akhiezer.)
M., Gostekhizdat, 1st edition, 1948; 2nd edition, 1950, 416 pp.
\item Effect of impurities on the thermodynamical properties of
velocity of second sound in He II.  ZhETF {\bf 19} (1949) 42 (SW,
{\bf 1}, 20).
\item A remark on scattering of particles with vanishing energy.
ZhETF {\bf 18} (1948) 1146 (SW, {\bf 1}, 31).
\item Lifetime of slow positrons.  ZhETF {\bf 19} (1949)
183 (SW, {\bf 2}, 37).
\item Radiation in the collision of fast neutrons with protons.
(With I.M.~Shmushkevich.) Dokl. AN SSSR {\bf 64} (1949) 499 (SW,
{\bf 3}, 80).
\item On the determination of non-electromagnetic interaction
between electrons and neutrons. (With A.I.~Akhiezer.)  ZhETF {\bf
19} (1949) 558 (SW, {\bf 1}, 32).
\item On the $\beta$-decay of neutron. (With V.B.~Berestetsky.)
 ZhETF {\bf 19} (1949) 756 (SW, {\bf 2}, 67).
\item Electromagnetic radiation caused by exchange forces. (With
I.M.~Shmushkevich.) Dokl. AN SSSR {\bf 70} (1950) 33 (SW, {\bf 3},
81).
\item On the theory of liquid $~^3$He.  ZhETF {\bf 20} (1950)
919 (SW, {\bf 1}, 21).
\item Exchange collisions of fast nucleons with deuterons. I.
ZhETF {\bf 21} (1951) 1113 (SW, {\bf 3}, 82).
\item Exchange collisions of fast nucleons with deuterons. Dokl.
AN SSSR {\bf 78} (1951) 249 (SW, {\bf 3}, 83).
\item On conversion of charged $\pi$-meson into a neutral meson in
collisions with proton and deuteron. (With V.B.~Berestetsky.)
Dokl. AN SSSR {\bf 77} (1951) 803;  ZhETF {\bf 21} (1951) 1333
(SW, {\bf 3}, 84).
\item On the theory of production of many particles in a single
act. Dokl. AN SSSR {\bf 78} (1951) 889 (SW, {\bf 3}, 105).
\item Capture of $\pi$-particle in a deuteron. Dokl. AN SSSR {\bf
80} (1951) 47 (SW, {\bf 3}, 86).
\item On collisions of $\pi$-mesons with deuterons. (With
V.B.~Berestetsky.) Dokl. AN SSSR {\bf 81} (1951) 1019 (SW, {\bf
3}, 88).
\item The conduction of heat in a fully ionized gas at high
temperatures. (With V.B.~Berestetsky and B.L.~Ioffe) 1951 (SW,
{\bf 1}, 345).

\begin{center}

1952 -- 1956

\end{center}

\item On the theory of capture of $\pi$-particles by deuteron.
ZhETF {\bf 22} (1952) 129 (SW, {\bf 3}, 85).
\item Exchange collisions of flat nucleons with deuterons.
ZhETF {\bf 22} (1952) 624 (SW, {\bf 3}, 87).
\item On electrons emitted in the process of capture of
$\mu$-mesons on atomic levels. (With B.L.~Ioffe.) ZhETF {\bf 23}
(1952) 123 (SW, {\bf 2}, 38).
\item On the spectrum of $\mu$-mesohydrogen. (With A.D.~Galanin.)
Dokl. AN SSSR {\bf 86} (1952) 251 (SW, {\bf 2}, 39).
\item On the emission of high energy $\gamma$-quanta in collisions
of fast neutrons with protons. (With I.M.~Shmushkevich.) Dokl. AN
SSSR {\bf 87} (1952) 385 (SW, {\bf 3}, 89).
\item On the paramagnetic dispersion. (With A.I.~Akhiezer.) Dokl.
AN SSSR {\bf 87} (1952) 917 (SW, {\bf 1}, 17).
\item Emission of $\gamma$-quanta in collisions of fast
$\pi$-mesons with nucleons. (With L.D.~Landau.) ZhETF {\bf 24}
(1953) 505; CERN Symp. {\bf 2} (1956) 159 (SW, {\bf 3}, 97).
\item On external (diffraction) generation of particles in nuclear
collisions. (With E.L.~Feinberg.) Dokl. AN SSSR {\bf 93} (1953)
439 (SW, {\bf 3}, 98).
\item Limits of applicability of the theory of Bremsstrahlung by
electrons and production of pairs at high energies. (With
L.D.~Landau.) Dokl. AN SSSR {\bf 92} (1953) 535 (SW, {\bf 2}, 51).
\item Electron-avalanche processes at superhigh energies. (With
L.D.~Landau.) Dokl. AN SSSR {\bf 92} (1953) 735 (SW, {\bf 2}, 5).
\item Emission of photon in the capture of a fast proton by
nucleus. (With A.I.~Akhiezer.) Dokl. AN SSSR {\bf 94} (1954) 821
(SW, {\bf 3}, 99).
\item Renormalization of mass and charge in covariant equations of
quantum field theory. (With A.D.~Galanin and B.L.~Ioffe.) Dokl. AN
SSSR {\bf 98} (1954) 361 (SW, {\bf 2}, 57).
\item Semiphenomenological theory of production of $\pi$-meson
pairs by high energy $\gamma$-quanta. Dokl. AN SSSR {\bf 96}
(1954) 265 (SW, {\bf 3}, 100).
\item Production of $\pi$-meson pairs by $\gamma$-quanta in heavy
nuclei.  Dokl. AN SSSR {\bf 96} (1954) 481 (SW, {\bf 3}, 101).
\item On asymptotics of nucleon Green function in pseudoscalar
theory with weak interaction. (With A.D.~Galanin and B.L.~Ioffe.)
ZhETF {\bf 29} (1955) 51 (SW, {\bf 2}, 58).
\item Generalization of Ward theorem for finite
wave-length of light in the case of particles of spin 0.  Dokl. AN
SSSR {\bf 100} (1955) 41 (SW, {\bf 2}, 59).
\item On the point interaction in quantum electrodynamics. (With
L.D.~Landau.)  Dokl. AN SSSR {\bf 102} (1955) 489 (SW, {\bf 2},
60).
\item Vanishing of renormalized charge in quantum electrodynamics.
 Dokl. AN SSSR {\bf 103} (1955) 1005 (SW, {\bf 2}, 61).
\item On  renormalization of meson charge in pseudoscalar theory
with pseudoscalar coupling.  Dokl. AN SSSR {\bf 104} (1955) 51
(SW, {\bf 2}, 62).
\item On vanishing renormalized meson charge in pseudoscalar
theory with pseudoscalar coupling.  Dokl. AN SSSR {\bf 105} (1955)
461 (SW, {\bf 2}, 63).
\item Solution of equations of pseudoscalar meson theory with
pseudoscalar coupling. ZhETF {\bf 29} (1955) 869 (SW, {\bf 2},
64).
\item Creation of $\mu$-meson pair in positron annihilation.
(With V.B.~Berestetsky.) ZhETF {\bf 29} (1955) 864 (SW, {\bf 2},
40).
\item On emission of $\gamma$-quanta in absorption of fast protons
by nuclei. (With A.I.~Akhiezer.) ZhETF {\bf 30} (1956) 201 (SW,
{\bf 3}, 102).
\item Vanishing of renormalized charge in electrodynamics and
meson theory. Nuovo Cim. {\bf 3} (1956) 1186 (SW, {\bf 2}, 65).
\item Isotopic invariance and scattering of antinucleons by
nucleons. ZhETF {\bf 30} (1956) 423 (SW, {\bf 3}, 106).
\item Isotopic invariance and cross section of interaction of high
energy $\pi$-mesons and nucleons with nucleons. (With L.B.~Okun.)
ZhETF {\bf 30} (1956) 424 (SW, {\bf 3}, 107).
\item Theory of resonance absorption in heterogeneous systems.
(With I.I.~Gurevich.) International Conference on peaceful use of
atomic energy. Geneva, 1955. M.: Publishing house of Academy of
Sciences of USSR, 1955, p.557 (SW, {\bf 1}, 26).
\item Vanishing of renormalized charge in theories with point
interaction. (With V.V.~Sudakov and K.A.~Ter-Martirosyan.) Phys.
Rev. {\bf 103} (1956) 784 (SW, {\bf 2}, 66).
\item Correlation effects in K-meson capture. (With
V.B.~Berestetsky.) ZhETF {\bf 31} (1956) 350 (SW, {\bf 2}, 68).
\item Dispersion relations for the scattering of $\pi$-mesons on
deuterons. (With B.L.~Ioffe and A.P.~Rudik.) ZhETF {\bf 31} (1956)
712 (SW, {\bf 3}, 90).
\item On the number of different types of K-mesons. (With
B.L.~Ioffe and L.B.~Okun.) Nucl. Phys. {\bf 2} (1956/1957) 277
(SW, {\bf 2}, 69).
\item Note on the number of different types of K-mesons. Nucl.
Phys. {\bf 2} (1956/1957) 281 (SW, {\bf 2}, 70).
\item Inelastic diffraction processes at high energies. (With
E.L.~Feinberg.) Nuovo Cim. Suppl. {\bf 3} No.4 (1956) 652 (SW,
{\bf 3}, 103).

\begin{center}

1957 -- 1961

\end{center}

\item On the possible transition dipole moment of
$\Lambda$-particles. (With B.L.~Ioffe.) Dokl. AN SSSR {\bf 113}
(1957) 1251 (SW, {\bf 2}, 71).
\item Equality of total cross sections of interaction of nucleons
and antinucleons at high energies. ZhETF {\bf 34} (1958) 725 (SW,
{\bf 3}, 108).
\item On interaction of $\Xi$-hyperons with nucleons and light
nuclei. (With L.B.~Okun and I.M.~Shmushkevich.) ZhETF {\bf 34}
(1958) 1246 (SW, {\bf 2}, 72).
\item On determination of parity of K-mesons. (With L.B.~Okun.)
ZhETF {\bf 34} (1958) 997 (SW, {\bf 2}, 73).
\item On the possibility to formulate the theory of strongly
interacting fermions. (With A.A.~Abrikosov, A.D.~Galanin,
L.P.~Gorkov, L.D.~Landau and K.A.~Ter-Martirosyan.) Phys. Rev.
{\bf 111} (1958) 321.
\item Diffraction effects in collisions of fast particles with
nuclei. (With A.I.~Akhiezer.) UFN {\bf 55} (1958) 593 (SW, {\bf
3}, 104).
\item Green functions in meson theories. (With A.A.~Abrikosov,
A.D.Galanin, B.L.~Ioffe and I.M.~Khalatnikov.) Nuovo Cim. {\bf 8}
(1958) 782.
\item On the stability of Fermi liquid.  ZhETF {\bf 35} (1958)
524 (SW, {\bf 1}, 22).
\item Isotopic effect in the residual electric resistance of
metals.  ZhETF {\bf 35} (1958) 992 (SW, {\bf 1}, 5).
\item On interaction between the conducting electrons in
ferromagnets. (With A.I.~Akhiezer.) ZhETF {\bf 36} (1959) 859 (SW,
{\bf 1}, 6).
\item On peripheral interactions of elementary particles.
(With L.B.~Okun.) ZhETF {\bf 36} (1959) 300; Nucl. Phys. {\bf 10}
(1959) 492 (SW, {\bf 3}, 91).
\item $\beta$-interaction and the nucleon form-factor.
(With V.B.~Berestetsky.) ZhETF {\bf 36} (1959) 1321 (SW, {\bf 2},
74).
\item On collision of nucleons at high angular momenta.
(With A.D.~Galanin, A.F.Grashin and B.L.~Ioffe.) ZhETF {\bf 37}
(1959) 1663; Nucl. Phys. {\bf 17} (1960) 181 (SW, {\bf 3}, 92).
\item Nucleon-nucleon scattering in two-meson approximation at
high angular momenta. (With A.D.~Galanin, A.F.~Grashin and
B.L.~Ioffe.) ZhETF {\bf 38} (1960) 475 (SW, {\bf 3}, 93).
\item On the limits of applicability of the transition radiation
theory. (With G.M.~Garibyan.) ZhETF {\bf 37} (1959) 1828 (SW, {\bf
2}, 53).
\item On asymptotic dependence of cross sections at high energies.
(With V.B.~Berestetsky.) ZhETF {\bf 39} (1960) 1078; Nucl. Phys.
{\bf 22} (1961) 629 (SW, {\bf 3}, 109).
\item On the theory of scattering of slow neutrons in a Fermi
liquid. (With A.I.~Akhiezer and I.A.~Akhiezer.) ZhETF {\bf 41}
(1961) 644 (SW, {\bf 1}, 23).
\item On processes of interaction of $\gamma$-quanta with unstable
particles. (With I.M.~Shmushkevich.) Nucl. Phys. {\bf 23} (1961)
452 (SW, {\bf 3}, 94).
\item Phase analysis of $pp$-scattering at energy 95 MeV.
(With V.A.~Borovikov, I.M.~Gelfand and A.F.~Grashin.) ZhETF {\bf
40} (1961) 1106.
\item On electromagnetic interaction of neutral vector meson.
(With I.Yu.~Kobzarev and L.B.~Okun.) ZhETF {\bf 41} (1961) 495
(SW, {\bf 2}, 75).
\item On production of high energy beams of $\pi$-mesons.
(With Yu.P.~Nikitin and I.M.Shmushkevich.) ZhETF {\bf 41} (1961)
963 (SW, {\bf 3}, 95).

\begin{center}

1962 -- 1966

\end{center}

\item Asymptotic behaviour of processes of annihilation and
elastic scattering at high energies. (With V.N.~Gribov.) Nucl.
Phys. {\bf 33} (1962) 516; Intern. Conf. Theor. Aspects Very High
Energy Phenomena. CERN, 1961, p.376; Preprint ITEP-61-15, M., 1961
(SW, {\bf 3}, 110).
\item Complex angular momenta and relations between cross sections
of various processes at high energies. (With V.N.~Gribov.) ZhETF
{\bf 42} (1962) 1141; Phys. Rev. Lett. {\bf 8} (1962) 343 (SW,
{\bf 3}, 111).
\item On certain properties of elastic scattering amplitudes at
high energies. (With V.N.~Gribov.) ZhETF {\bf 43} (1962) 308;
Nucl. Phys. {\bf 38} (1962) 516 (SW, {\bf 3}, 112).
\item Certain corollaries of the hypothesis of the moving poles
for the processes at high energies (With V.N.~Gribov, B.L.~Ioffe
and A.P.~Rudik.) ZhETF {\bf 42} (1962) 1419 (SW, {\bf 3}, 113).
\item On the scattering of slow neutrons in Fermi liquid.
(With A.I.~Akhiezer.) Nucl. Phys. {\bf 40} (1963) 139 (SW, {\bf
1}, 24).
\item Spin structure of meson-nucleon and nucleon-nucleon
scattering at high energies. (With V.N.~Gribov.) ZhETF {\bf 42}
(1962) 1682; Phys. Rev. Lett. {\bf 8} (1962) 412 (SW, {\bf 3},
114).
\item Regge poles and Landau singularities. (With V.N.~Gribov.) ZhETF
{\bf 43} (1962) 1970; Phys. Rev. Lett. {\bf 9} (1962) 238; Proc.
Intern. Conf. High Energy Phys. Geneva, 1962, p.543 (SW, {\bf 3},
115).
\item Limitation on the rate of decrease of amplitudes in various
processes. (With V.N.~Gribov.) ZhETF {\bf 43} (1962) 1556; Phys.
Lett. {\bf 2} (1962) 239; Proc. Intern. Conf. High Energy Phys.
Geneva, 1962, p.522 (SW, {\bf 3}, 116).
\item Om processes determined by fermionic Regge poles.
(With V.N.~Gribov and L.B.~Okun.) ZhETF {\bf 45} (1963) 114 (SW,
{\bf 3}, 117).
\item Singularities  of partial waves close to $j=1$ and behaviour
of elastic scattering amplitude at high energies. (With
V.N.~Gribov and K.A.~Ter-Martirosyan.) Phys. Lett. {\bf 9} (1964)
269 (SW, {\bf 3}, 118). A remark to this paper: Phys. Lett. {\bf
12} (1964) 153 (SW, {\bf 3}, 340).
\item Certain corollaries  of unitary symmetry for processes
involving $\omega$-, $\varphi$- and $f^0$-mesons. (With B.L.~Ioffe
and I.Yu.~Kobzarev.) ZhETF {\bf 48} (1965) 375 (SW, {\bf 2}, 76).
\item Moving branching points in the $j$-plane and Regge conditions
of unitarity. (With V.N.~Gribov and K.A.~Ter-Martirosyan.) Yad.
Fiz. {\bf 2} (1965) 361; Phys. Rev. {\bf 139} (1965) 184; Problems
in elementary particle physics. Erevan, 1964, p.167 (SW, {\bf 3},
119).
\item Structure of $j$-plane near $j=1$ and diffraction scattering
at high energies. (With V.N.~Gribov and K.A.~Ter-Martirosyan.)
Preprint ITEP No.238, M., 1964.
\item ``Shadow universe'' and neutrino experiment. (With
L.B.~Okun.) Pisma v ZhETF {\bf 1} (1965) 28; Phys. Lett. {\bf 16}
(1965) 338 (SW, {\bf 2}, 77).
\item Electromagnetic mass differences of baryons and
SU(6)-symmetry. (With A.D.~Dolgov, L.B.~Okun and V.V.~Solovyev.)
Yad. Fiz. {\bf 1} (1965) 730; Phys. Lett. {\bf 15} (1965) 84 (SW,
{\bf 2}, 78).
\item At which distances  does the high energy interaction take
place? (With V.N.~Gribov and B.L.~Ioffe.) Yad. Fiz. {\bf 2} (1965)
768 (SW, {\bf 3}, 120).
\item On the possibility of experimental detection of mirror
particles. (With I.Yu.~Kobzarev and L.B.~Okun.) Yad. Fiz. {\bf 3}
(1966) 1154 (SW, {\bf 2}, 79).
\item On total cross section of annihilation of electron-positron
pairs into hadrons at high energies. (With V.N.~Gribov and
B.L.~Ioffe.) Yad. Fiz. {\bf 6} (1967) 587 (SW, {\bf 3}, 121).
\item The formula by Orear  as a consequence of branching points
in the $j$-plane (SW, {\bf 3}, 122).
\item Limitation on the rate of increase of cross sections for
weak interactions.  Yad. Fiz. {\bf 11} (1970) 852 (SW, {\bf 3},
123).
\end{enumerate}

\end{document}